# On the Potential of Large Ring Lasers

G. E. Stedman[1,1], R. B. Hurst[1] and K. U. Schreiber[3]

[1]Department of Physics University of Canterbury, Private Bag 4800, Christchurch 2, New Zealand
[3] Forschungseinrichtung Satellitengeodäsie der Technischen Universität München, Fundamentalstation Wettzell D-93444 Bad Kötzting, Germany

## Abstract

We describe a new ring laser with area $A = 833$ m$^2$ and update performance statistics for several such machines. Anandan & Chaio 1982 judged ring lasers inferior to matter interferometers as possible detectors of gravitational waves. However, we note that geophysically interesting results have been obtained from large ring lasers and that there is still a lot of room for improvements.

## 1      Introduction

In their extended footnote #11 Anandan and Chaio [1] compared ring laser interferometers and matter interferometers as detectors of gravitational waves. They write in part …[in the case of a Sagnac optical interferometer], in equation (7) of [1] ) "the mass $m$ must be replaced by $(\hbar\omega/c^2)$, where $\omega$ is the frequency of light relative to the apparatus at the point of interference. For optical frequencies $\hbar\omega/(mc^2)$ $\sim 10^{-9}$, so that this device [ring interferometer] would be much less sensitive than the matter interferometer." Indeed it has long been expected from such arguments that matter interferometers will eventually prove superior to all other ring interferometry [2]. However, in the years after Anandan & Chaio's paper [1] and before Josephson effects were realised in super-fluid junctions, a revolution in mirror design has meant that reflectivities rose until they are now > 99.999% so that cavity quality factors and finesses have risen to values of $2 \cdot 10^{13}$ and $10^5$ respectively [3][4]. Since [4] our team now has constructed a rectangular laser gyroscope **UG2** of 21m x 39.7m with area $A = 833.7$m$^2$ and perimeter $P = 121.4$ m as an upgrade since November 2005 of an earlier and smaller laser, **UG1.** In addition a new generation of super-mirrors has been installed in this machine, also in an earlier machine **C-II.** It is therefore time to update Table 1 of an earlier study [4] and this is done in the present paper. We also report new results for beam powers, sizes, and ring-down times and their consequences for possible sensitivity. For example, we use the recently measured ringdown time of 1 ms for **G**, 0.1 ms for **PR1** and 400 μs for **UG2**. The figure for **UG2** is lower than expected from such a large cavity and reasons for this are still obscure and under investigation. It seems clear from the visible haloes scattered from some mirrors that aberrations arise from the figuring of the super-mirror surfaces, another factor may be relatively high absorption in the ULE substrates used.

The report in Section 2 on these machines indicates capabilities of optical Sagnac interferometry well beyond the level that could have been conceived by Anandan & Chaio in 1982 [1]. In particular there is now evidence that relatively large ring lasers are practical devices with the potential to reveal new physical effects and in particular geophysical effects, such as:

---

[1] geoffrey.stedman@canterbury.ac.nz

- Earth tides

    In [5] it was explained how **C-II** is capable of observing solid Earth tides at the frequencies 22.304 µHz, caused by tilts from crustal deformation and ocean loading from lunar attraction. These tilts with amplitudes of about 80 nrad alter the projection of Earth's rotation on the ring. The contributions of sea tides and cavity corrections on the local cavern tilts were also considered in detail.

- Lunar nutations

    In [6] it was described how **C-II, UG1** and **G** have all observed daily motions of the Earth's rotation axis of a magnitude of 45 cm at the poles driven by the effects of lunar gravity on an in-homogeneous mass distribution of the Earth. The frequencies involved are 10.775 and 11.613 $\mu$Hz.

- Earthquake-induced ground rotations

    This field of application is discussed e.g. [7] [8] and more widely in [27].

As opposed to the Michelson type of optical interferometers currently favored in gravitational wave detection, all our machines are active He-Ne ring interferometers which detect the Sagnac frequency signal arising from Earth's rotation and perturbations thereof, and are in principle also sensitive to external rotational excitations such as gravitational wave induced fluctuations in any component of the Riemann tensor which varies the speed of light in the ring interferometer. In particular, in the components $g_{0i}$ of the metric tensor, some discussions already exist on the effect of gravitational radiation on a gyroscope including [9]-[14].

We are not attempting here a full proposal for an optical gravitational wave interferometer. This would require a study of including orders of magnitude of possible source signals, antenna design and polarization, optimum practical size and engineering, and the correspondingly relevant Riemann tensor components, effects of various strains and rotations, and associated noise in the geophysical environment. Our aim here is (1) to introduce **UG2** (2) to raise the question as to whether ring lasers still merit the negative conclusions made by Anandan & Chaio in 1982 [1]. In our opinion they do not.

We accept that many orders of magnitude improvement in their performance will be needed. When we use the de-Broglie relation $\lambda=\hbar/mv$ in the standard Sagnac formula for a phase difference, $\Delta\varphi = 8\pi \boldsymbol{A} \cdot \boldsymbol{\Omega}/\lambda v$ gives $8\pi m / (\hbar \boldsymbol{A} \cdot \boldsymbol{\Omega})$ and the $m/\hbar$ factor is dominant. Such arguments remain valid in principle and certainly indicate that matter wave interferometry has potential to outclass optical interferometry [2]. However, that is not to say that we endorse all such claims in the context of ring interferometry [15] [16]. Apart from such theoretical considerations one also has to keep in mind that these have to be translated into an apparatus, where experimental challenges may be hard to overcome. The relative simplicity of optical laser gyro designs is an important factor for their success.

## 2    Present ring lasers

The gyroscopes discussed here have the geometry given in Table 1, which includes relevant references for each laser. The most recent performance data is given in Table 2. These tables update the table in [4]. The machines whose results we discuss include the climax of two decades of development of various rectangular He-Ne gas ring lasers first at University of Canterbury, Christchurch, New Zealand, whose underground laboratory at Cashmere has latitude 43.57694 S and longitude 172.622 E and another underground laboratory at Wettzell in Bavaria, Germany, with latitude 49.14441 N and longitude 12.878 E.

**C-II** has a monolithic construction out of a 18 cm thick Zerodur slab, was manufactured by Zeiss and was recently upgraded (December 2005) with 4 modern super mirrors from REO with nominal losses of 4 parts per billion per mirror.

**G0** is a vertical stainless steel laser structure mounted on a concrete wall of the Cashmere Cavern [19] and showed that the previously un-attempted size of **G** was possible.

**UG1** also used stainless steel vacuum tubing anchored into the rock floor of the cavern and following [20] the larger rings use a higher gas pressure than normal to reduce mode competition 25:1 He:Ne at 5 or even 40:1 He:Ne at 8 mbar.

**UG2** is rectangular in shape and replaces **UG1.** It has an enlarged footprint and goes around the entire Cashmere Cavern area.

**PR1** (1.6m square and with new generation super-mirrors**)** in the Physics Department building, University of Canterbury, monitors seismic rotations of buildings.

The **GEOsensor (GEOs)** is a simplified ring laser, specifically build for seismic studies [37], where short term stability is required. It is identical to **PR1** and located at the Pinon Flat Seismological Observatory, (CA) of the Scripps Institution of Oceanography at UCSD.

**G** is a high precision nearly monolithic ring laser machine in the German laboratory and has measured rotation rates as small as $10^{-10}$ rad/s/$\sqrt{Hz}$ [17][18]. **G** has a Zerodur slab 25 cm thick and 4.2 m in diameter, weighing approximately 9 t.

This range of instruments has discredited fears that untamable mechanical instability and also fundamental quantum noise limitations would make significantly large rings impossible [16][7]. However we have to note, that a much more elaborate mechanical sensor design is required to fully utilize the potential of these larger rings. The currently explored structures cannot make full use of their enhanced sensor resolution.

Two proposals have been suggested for still larger such machines [4]. One ring laser, **UG100**, would be 100 m square with a radical different cavity design and the other **MCR** for **M**ichelson **C**entennial **R**ing is proposed to update the classic work of Michelson, Gale and Pearson in 1925 [21] to the laser era. By using a laser Sagnac interferometer, as opposed to the purely classical interferometer, built by Michelson and Gale $A = 207800$ m$^2$ $P = 1904$ m, such a proposal would seem a fitting objective with date of completion say 2025, the anniversary of Michelson, Gale and Pearson's work. We know of no fundamental obstacle to the viability of either of these proposals. Neither of our present laboratories could house such proposals, either of which would be a worthy step towards a ring for gravity wave detection, at least from a point of view of enlarged scaling factor. At some time in the future it may well be important to have an alternative measurement concept available when it comes to investigating properties of gravitational waves.

The viability of ring lasers as gravitational wave detectors in a sufficiently large system like say **UG100** or **MCR** might be gauged in part from Allan variance plots which certainly confirm that microseismic noise and other geophysical signals are significant in the background; this is not reduced with the use of a larger ring. We would anticipate that in any full instrument design, one would be required to move to a DC measurement system if only adequate stability were possible. A shift of frequency towards 1 kHz say and decoupling the ring interferometer from the Earth background would surely be desirable, however this may be in conflict with basic operational parameters of a ring laser. This would undoubtedly require a dramatic change in design, e.g. with beam compression, and gas confining enclosures, something which is not used in any of the lasers of Tables 1 and 2. Of course, all gravitational wave detectors are deeply affected by quantum noise limits [22]. The question at stake is whether ring lasers can be improved to reach the level of performance necessary to detect gravitational waves. We suggest that, in view of recent results with ring lasers, this topic is well worth a fuller study; no doubt many questions will be resolved only by the building of a prototype e.g. **UG100** extendable to **MCR.** We have already noted [4] prior to the construction of **UG2** that we expect significantly better performance due to an enlarged scale factor at sizes of $A = 10^3 - 10^5$ m$^2$. In the absence of external perturbations the Allan deviations of **UG2, UG100** and **MCR** at 1000 s relative to the Sagnac frequency contribution from the Earth's rotation rate were estimated in [4] to be 0.062, 0.16 and 0.008.

One measure of the performance of a ring is its Allan variance. These are plotted in fig. 1 relative to the Sagnac frequencies, for all the recent lasers **C-II**, **G0, G, UG1** and **UG2**. Fig.1 shows that **UG2** has in practice achieved a normalized value of the order of $10^{-6}$ at 2000 s, not as good as the much smaller ring **UG1**. This is a good demonstration for the growing difficulties with mechanical stability of larger constructions. All this of course is still well short of the $10^{-20}$ regime needed for gravity wave detection. Photon quantum noise in the lasing gas contributes to the lower bounds in this plot at low times [7], experience with LISA and LIGO illustrate how such obstacles may be overcome with laser systems, given sufficient ingenuity and effort. If large rings can reach a performance of significance it will also depend in part on the frequency window of the gravitational wave spectrum of interest; for example in the space borne mission LISA it is $10^{-4}$ -1 Hz [23].

The Allan deviation is dominated by the drifts of known origin in **UG2**. At the present times studies are being undertaken of their correlations with thermo-elastic cavern deformation, beam position drifts and laser gain degradation through out-gassing. Such performance data together with a full analysis of all relevant noise sources are vital components of a full appraisal of the potential of ring laser detectors. Immediately after the mirror upgrade of November 2005 in **C-II** and **UG2** it was noted that **C-II** Sagnac frequency was stable to ½ mHz per day, with an angular sensitivity of 0.6333 nrad, while **UG2** had a corresponding sensitivity of 1 µrad. On the other side the monolithic **G** routinely resolves 1 prad/s at 2000 seconds of integration. The discussion here has been centered exclusively on one aspect of gyroscopy, namely sensor sensitivity. While sensor resolution is important to actually resolve relativistic quantities, instrumental up-scaling of our gyroscopes alone is not enough for clear GW detection. Such sensitivities along with the estimate of Table 2, obtained using the method of [4] with the ring-down times, give some hope that the regime of interest for general relativity is not necessarily inaccessible to ring laser interferometers. Which may be compared with particular values for detection of gravitational radiation detection in the relatively low frequency detection bands, provided this is not prohibited by perturbations from Earth generated noise. It has already been argued by MIGO [24] that a natural advantage applies to matter interferometers at higher (kHz) frequency bands. Other electromagnetic detectors have been claimed to have promise for relatively high frequency [25] [26].

## 3    Prerequisites for the Detection of Relativistic Effects

The interferogram of a ring laser gyroscope contains contributions of several different processes and this needs careful considerations when applications such as gravitational wave detections are considered for a rotation sensing device. Some of these effects are of general nature and apply to all gyroscopes, while others are specific to ring lasers. However, other techniques like superfluid helium gyroscopes and atom interferometry will also suffer from technique inherent side-effects, which may be easier or more difficult to overcome. Here we attempt to outline the known challenges for ring laser gyroscopes.

**Scale factor:** One way of making ring laser gyroscopes more sensitive can be achieved by up-scaling the geometrical size of the instrument. By going from **C-II** (1 m$^2$) to **UG2** (833 m$^2$) we have followed this path. As a consequence of this effort it was not possible to construct a monolithic instrument for sizes larger than 16 m$^2$. The heterolithic design of our larger rings **UG1** and **UG2**, made from steel pipes and concrete pedestals attached to the floor of the Cashmere Cavern however can not provide the necessary mechanical rigidity required for studies of fundamental physics as shown in fig. 1. The comparison of the achieved sensor stability expressed in terms of the Allan deviation, shows the lowest values for the largest monolithically realized ring laser and reduces in quality considerably for the heterolithic rings. With the tilts from solid Earth tides [5] also comes Earth strain, which changes the perimeter and area and thus the scale factor of the rings **UG1** and **UG2**. For the case of a square ring and the absence of shear forces these scale factor variations are compensated by the corresponding shift of the optical frequency of the ring cavity [18] as long as the longitudinal mode index of the oscillating laser mode stays the same. On the other hand there are significant variations of the scale factor from thermo-elastic deformations of the cave under the variable influence of the atmospheric pressure load.

**Orientation**: Until today all relevant geophysical signals that our ring lasers detected were small periodic signals with periods of several seconds (seismic signals), half a day (solid Earth tides) and the sidereal day (polar motion). In particular the latter 2 signals only became observable because of a change of the projection of the normal vector of the ring lasers onto the Earth rotation vector at amplitudes of around 80 nrad. Much larger geophysical signals such as the Chandler wobble (period ≈ 435 days) are still outside of our measurement window, because of their quasi-static nature. The instrumental drift on a per day basis is much larger than the amplitude of this oscillation. Tilts and strain-induced rotations resulting from crustal deformations caused by variations of atmospheric loading have not yet been identified, because they are not strictly periodic in nature. Finally, we are not yet in the position to determine the contributions of the variations of the ocean angular momentum and the atmospheric angular momentum to the Earth rotation rate. Both, the instrumental sensitivity and the sensor stability of the best performing ring laser (**G**) are still about one order of magnitude short of this goal.

**Sensor drift:** Ring lasers are actively lasing interferometers and as such contain a gain medium within the cavity, which is subject to slow degradation with time. In particular our large ring lasers are using a mixture of Helium and Neon gas together with r.f. excited plasma discharges. Due to out-gassing from the enclosure of the cavity, one can see a gradual decrease in laser gain, which in turn causes a slow drift in the Sagnac frequency at the level of up to 10 parts per million per day. This varies from one instrument to the other, but certainly places a severe limitation on the measurability of variations in rotation rate in all of our instruments. Unlike gravitational wave antennas our ring lasers run on low circulating beam powers (5-10 mW). Therefore quantum noise and phase jitter generate another severe limitation for the detection of signals of fundamental physics. This time the detection of higher frequency signals is compromised, rather than low frequency signals as discussed so far.

**Background Noise (Earth):** A ring laser is a local self-contained sensor, which is not depending on an external reference frame. When attached to the Earth, we can eventually expect to monitor the Earth rotation in great detail, provided the link between the gyroscope and the body of the Earth is rigid enough. As mentioned above, the thermo-elastic deformation of the upper crust (also as a result of variations in atmospheric pressure) is certainly a limiting factor. Micro-seismic activity in the frequency band of 0.2 - 0.5 Hz for example is a well known limitation, which may affect a gyroscope both in orientation as well as in the rotation rate through instrumental dither. Such signals may also cause beam wander in heterolithic ring laser constructions, which in turn generate scale factor variations. While beam wander has been successfully measured over time intervals of several minutes up to several days, we may well find also contributions in the millisecond regime.

All these various problems remain to be overcome for any gravitational wave detector based on ring lasers to become feasible. Shot noise is a particular big challenge, because in current designs we depend on mono-mode operations. While it would be desirable to go to say 1 kHz or higher frequencies for the signal band of interest, this interferes with shot noise limitations, since there is the problem of detecting a very small phase change unambiguously out of a noisy frequency itself.

Our aim here is to note that since an area of 833 m$^2$ has been achieved for ring laser operation, there is no fundamental limit evident that would prohibit further progress. It is also worth noting that Michelson antennas and Sagnac antennas operate in the same sort of ball-park in terms of frequencies and amplitudes. The latter would provide rotational components from gravitational waves. Such components are discernable in the works of Chaio & Anandan[1], for example.

Such an independent form of measurements with Michelson and Sagnac antennas could be valuable. The relationship between rotation rate and translational components may be an important step in the verification of gravitational waves: Two different technologies extract different quantities, whose a priori relationship merits fuller study than that attempted here. A combined observation may be achieving more than would be possible with a single measurement technique. Just as in rotational seismology there is a strong formal link between earthquake induced rotations and translations via the transverse acceleration [27].

## 4    The status of matter wave Sagnac interferometer

In the spirit of Anandan and Chaio [1] we briefly outline recent work in matter interferometry. The range may be gauged here of work from the many links and references on [28] Packard's web site. Special mention may be made of atom or electron interferometers by Hasselbach [29] [30], the Kasevich group [31]-[33], and we note also the Varoquaux Group [34]. The effective area may be increased by combining matter interferometers with optical techniques [35]. The Kasevich group quotes sensitivities of $6.10^{-8}$ rad/s for 1s integrations, and few % of the Earth rotation signal $\Omega_E$.

## 5    Conclusions

Such sensitivities along with the estimate of Table 2, obtained using the method of [4] with the ring-down times, give some hope that the regime of interest for general relativity is not necessarily inaccessible to ring laser interferometers. Fig. 1 demonstrates data on stability at short time scales, which may be compared with particular values for detection of gravitational radiation in the relatively low frequency detection bands. It has already been argued by MIGO [24] that a natural advantage applies to matter interferometers at higher (kHz) frequency ranges. Other electromagnetic detectors have been claimed to have promise for relatively high frequency [25] [27].

Obviously an enormous amount of work still needs to be done on laser interferometer gravitational wave detection, which is not discussed here. One review is [36]. We are not the first to suggest that Earth-bound interferometers of a relatively modest size could be of interest in gravitational wave detection e.g. [26].

The quality of the results in geophysics (section 1) with various ring lasers described in section 2 encourage larger proposals, which might be of a scale pointing towards gravitational wave detection, but neither of the two proposals so far proposed, **UG100** and **MCR,** are within the current resources of the German-New Zealand collaboration.

**Acknowledgements** GES thanks P.H. Butler for encouragement and D.I. Stedman for help in editing. U. Schreiber thanks M. Schneider and K. Danzmann for valuable discussions.

**Table 1**

The geometrical parameters for each of the lasers is based on a rectangle defined by four mirrors *A*, *B*, *C*, *D* where linear dimensions *a* and *b* give an area *A=ab* and a perimeter *P=2(a+b)*. The mirrors have radii of curvature $R_A = R_B$ and $R_C = R_D$. In the case of **C-II** from 1998-2005, $R_A$ = 6 m; $R_C = \infty$. Locations are Cashmere Cavern, Wettzell in Germany, ChCh=the Physics Department, University of Canterbury for **PR1**, and CA= Seismological Observatory Pinon Flat in California for **GEOs**,

| Laser | *a* | *b* | *A* | *P* | $R_A$ | $R_C$ | *Where* | date | refs |
|---|---|---|---|---|---|---|---|---|---|
| Units | m | *m* | m$^2$ | m | m | m | | | |
| **C-II** | 1 | 1 | 1 | 4 | 4 | *4* | Cashmere | 1997- | [17]-[19] |
| **G0** | 3.5 | 3.5 | 12.25 | 14 | 6 | 6 | Cashmere | 1998- | [19] |
| **G** | 4 | 4 | 16.0 | 16 | 4 | 4 | Wettzell | 2001- | [5][6] |
| **UG1** | 21.0 | 17.5 | 367.5 | 77 | 20 | 20 | Cashmere | 2003-5 | [20] |
| **UG2** | 21.0 | 39.7 | 833.7 | 121.4 | 20 | 70 | Cashmere | 2005- | - |
| **PR1** | 1.6 | 1.6 | 2.56 | 6.4 | 4 | 4 | ChCh | 2004- | - |
| **GEOs** | 1.6 | 1.6 | 2.56 | 6.4 | 4 | 4 | CA | 2005- | [8] |

**Table 2**

Parameters are as measured in August 2006 for beam power, lifetime etc. This is a significant update on the corresponding table 1 of [4]. Corresponding geometrical data is given in Table 1. $\delta f_E$ is the Sagnac frequency induced by Earth rotation of the ring from equation (2) of [7] $\delta f_E = 4\mathbf{A}\cdot\mathbf{\Omega}/\lambda P$ where $\mathbf{A}\cdot\mathbf{\Omega}$ is the dot product of the earth rotation vector $\mathbf{\Omega}$ on to the area vector $\mathbf{A}$ of each ring. All rings are horizontally mounted except for **G0** [19] and **PR1**. Observed values are used for the more recent lasers. The value $\delta f_E$ for **G** is corrected from that in [4] which used the wrong latitude. We tabulate figures from two new rings **PR1** and **GEOs**, two essentially identical smaller rings. As in [4], $F, Q, \tau$ are the cavity finesse, quality factor and ring-down time, and $S$ is the total beam power loss per mirror from the mirror scattering and $T$ the transmission power loss. A measurement of $T$ on recent mirrors of $T=0.24$ ppm is used. $T$ and $S$ are related by [4] $F = \pi/2S = cQ/f_0P$, $Q = 2\pi f_0 \tau$ where $f_0$ is the He-Ne laser frequency (474 THz), $p$ is the emergent beam power at each port, $\Omega_N$ is the angular sensitivity defined as in equation (27) of [7], with $\Omega_N = (cP/4AQ)(hf_0/P_0t)^{1/2}$ and t the observation time. $\mathrm{rad}_{1000}$ is the Allan deviation relative to $\delta f_E$ expected from quantum noise at 1000 s. $\mathrm{rad}_{1000} = (hf_0^3/2000P_0)^{1/2}/\delta f_E Q$. $d_{calc}$ and $d_{obs}$ are measured and predicted average beam diameters at the mirrors.

| Laser | $d_{calc}$ | $d_{obs}$ | $p$ | $\Omega_N$ | $\delta f_E$ | $\mathrm{rad}_{1000}$ |
|---|---|---|---|---|---|---|
| Units | mm | mm | nW | prad/s $\sqrt{\mathrm{Hz}}$ | kHz | ppb |
| **C-II** | 2-4 | 2.7 | 1 | 41 | 0.079 | 18.4 |
| **G0** | 3.1 | 5.5 | 1.8 | 77 | 0.287 | 33.2 |
| **G** | 1-3 | 3.7 | 3 | 0.9 | 0.349 | 0.37 |
| **UG1** | 30 | 8.25 | 10 | 0.4 | 1.513 | 0.18 |
| **UG2** | 2-3 | 13.4 | 10 | 0.2 | 2.177 | 0.01 |
| **PR1** | 2-3 | 1.05 | 5 | 10 | 0.16 | 5.2 |

**Figure 1:** Allan variances for **C-II**, **G0**, **G**, **UG1** and **UG2** normalized by their Sagnac frequencies. All the rings with a normalized Allan deviation of less than 1 ppm successfully showed geophysical signals from solid Earth tides and diurnal polar motion.

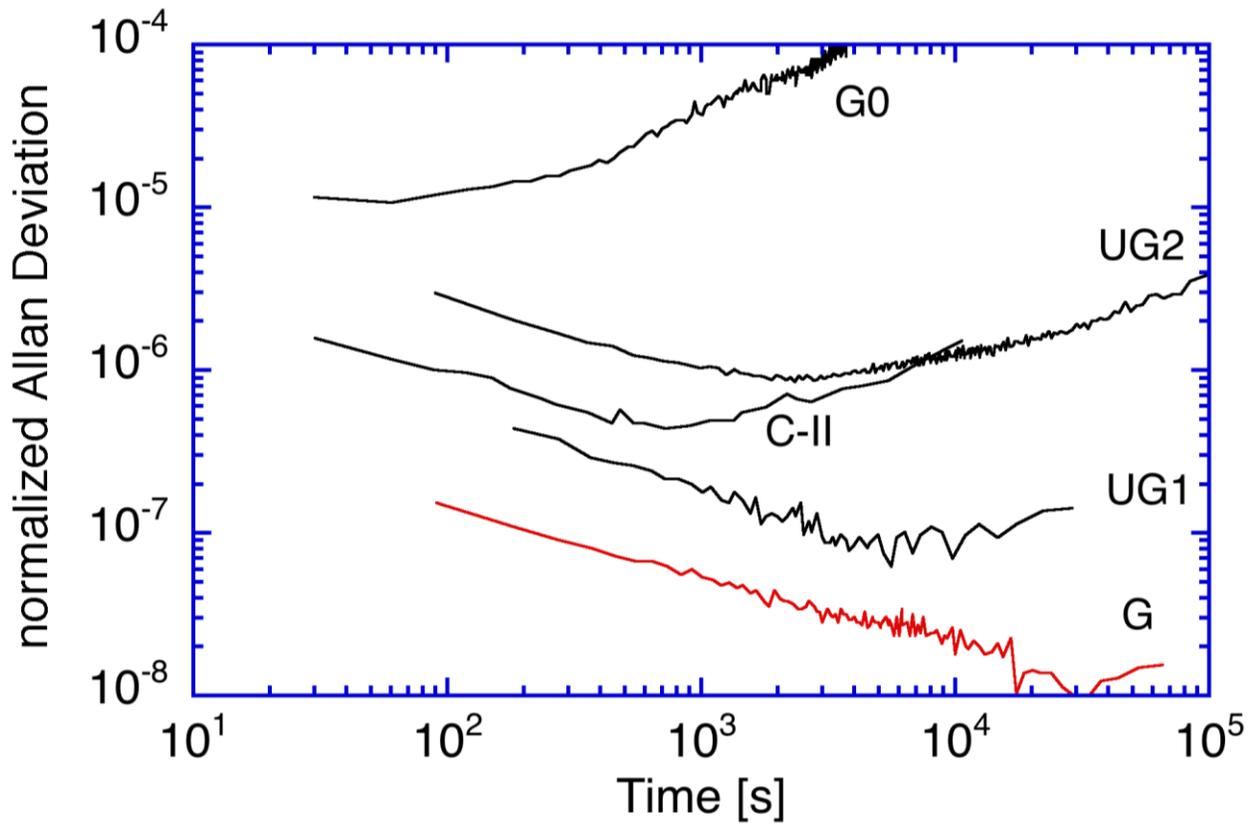